\documentclass[aps,prd,twocolumn,showpacs,nofootinbib,amsmath,amssymb,amsfonts,showkeys]{revtex4}

\usepackage{graphicx}
\usepackage{bm}

\begin{document}
\title{Distinguishability of scalar field models of dark energy \\ with time variable equation of state parameter}

\author{B. Novosyadlyj}
 \email{novos@astro.franko.lviv.ua}
\author{O. Sergijenko}
 \email{olka@astro.franko.lviv.ua}
\author{S. Apunevych}
 \email{apus@astro.franko.lviv.ua}
\affiliation{Astronomical Observatory of 
Ivan Franko National University of Lviv, Kyryla i Methodia str., 8, Lviv, 79005, Ukraine}
 
\date{\today}

\begin{abstract}
The possibility of distinguishing of scalar field models of dark energy with different Lagrangians and time variable equation of state parameter by available observational data is analyzed. The multicomponent cosmological models with the scalar field with either Klein-Gordon or Dirac-Born-Infeld Lagrangians as dark energy and the monotonic decreasing and increasing equation of state parameters are considered. It is concluded that scalar field models of dark energy with decreasing and increasing EoS parameters should be distinguishable at the accuracy level of forthcoming observational data. The Lagrangians of scalar fields could be distinguished by expected observational data (Planck, SDSS etc.) in the case of decreasing EoS parameter, but are practically indistinguishable in the case of increasing one.

\end{abstract}
\pacs{95.36.+x, 98.80.-k}
\keywords{cosmology: dark energy--scalar field--cosmic microwave background--large scale structure of the Universe--cosmological parameters}
\maketitle

\section{Introduction}
The observational data on Supernovae of type Ia light curves, CMB anisotropy and large scale structure of the Universe are well fitted by the cosmological models with scalar field as dark energy. In numerous papers (see e.g. the latest reviews \cite{rev1,rev2,rev3,rev4,rev5,rev6,rev7,rev8}, books \cite{tb1,tb2,tb3} and citing therein) the physical, cosmological and astrophysical aspects of dark energy are elucidated. But the main question ``which field from the large number of candidates is preferable in the light of obtained up to now observational data?'' remains without answer. The ``goodness of fit'' of many of them, which are quite different, for the same dataset is so close that the problem of principal distinguishability of scalar field models of dark energy arises. In the paper we analyze this problem for the subclass of minimally coupled dynamical scalar fields.\\
\indent In the fluid approach the minimal number of dark energy parameters required for description of the dynamical and clustering properties of a scalar field is three: the current density parameter $\Omega_{de}$, equation of state (EoS) parameter $w$ and effective sound speed $c_s^2$. They in the general case are functions of the scale factor $a$ (or redshift $z=a^{-1}-1$). The last function can be deduced from scalar field Lagrangian when it is given. The function $w(a)$ usually is defined ad hoc. At the current level of possibility of constraining the dark energy parameters by observations the simplest linear function $w(a)=w_0+w_a(1-a)$ \cite{wa1,wa2} is widely used. 
Two constants $w_0$ and $w_a$ (present values of $w$ and its first derivative with respect to the scale factor with opposite sign) are determined together with density parameter $\Omega_{de}$ and other cosmological parameters, the minimal set of which contains six ones: density parameter of baryons $\omega_{b}$, density parameter of cold dark matter $\omega_{cdm}$, Hubble constant $H_0$, spectral index of initial matter density power spectrum $n_s$ (scalar mode), amplitude of initial matter density power spectrum $A_s$ and reionization optical depth $\tau_{rei}$. The current observational data on Supernovae type Ia light curves, CMB anisotropy and large scale structure of the Universe allow the possibility to determine most of them with high accuracy. The exception is parameter $w_a$: its uncertainty is so large, that the character of variation of EoS parameter -- increasing or decreasing -- is not recognized (see, for example, \cite{Zhao2007,Komatsu2009,Komatsu2010}). \\
\indent In the paper \cite{Novosyadlyj2010} we have determined the parameters of scalar field models of dynamical dark energy with the other parametrization of EoS, which follows from assumption that the temporal derivative of dark energy pressure is proportional to the temporal derivative of its energy density. This parametrization involves two free parameter $w_0$ and $w_e$, which are the values of EoS parameter at current and early epochs respectively. The mean likelihood distribution obtained by the Markov Chain Monte Carlo (MCMC) method for the available dataset on CMB anisotropy, power spectrum of spatial inhomogeneities in distribution of galaxies and SN Ia photometric curves has two peaks with corresponding sets of best fitting parameters $\textbf{p}_i=(\Omega_{de},\,w_0,\,w_e,\,\omega_{b},\,\omega_{cdm},\,H_0,\,n_s,\,A_s,\,\tau_{rei}$), $i=1,\,2$ (Table \ref{tabl1}). The first of them corresponds to the scalar field model with decreasing EoS parameter, the second one to the scalar field model with increasing EoS parameter. Since the $-\log{L}$'s for both $\textbf{p}_1$ and  $\textbf{p}_2$ are close, we have concluded that these models are indistinguishable by the used dataset. This situation is the same for both classical and tachyonic scalar fields. Moreover, they are indistinguishable too. \\
\indent In this paper we analyze the quantitative differences of this four models of dark energy -- classical and tachyonic scalar fields, each with $\textbf{p}_1$ and $\textbf{p}_2$ parameter sets, and discuss the possibility of their distinguishing in the light of available and expected observational data. \\
\indent The paper is organized as follows. In Section \ref{models} we discuss the scalar field models of dark energy and determination of their parameters. Section \ref{bckgr} is devoted to the analysis of background dynamics in the  models with increasing and decreasing EoS parameters of scalar field and diferences of SN Ia light curves. In Section \ref{spectra} the differences of matter density and CMB temperature fluctuations power spectra of all models are analyzed and compared with corresponding observational uncertainties. Section \ref{conclusions} summarizes the results of analyses and presents the conclusions. 

\section{Models}\label{models}
We analyze the dynamics of expansion and the large scale structure formation of the spatially flat Universe filled with the non-relativistic particles (cold dark matter and baryons), relativistic ones (thermal electromagnetic radiation and massless neutrino) and minimally coupled scalar field with given Lagrangian. We apply
Einstein equations in the Friedmann-Robertson-Walker (FRW) metric for background dynamics and Einstein-Boltzmann system of equations in the synchronous gauge
for the evolution of linear perturbations. The line-element of FRW metric is $ds^2=c^2dt^2-a^2(t)d\sigma^2$, where $d\sigma^2$ is time-independent Euclidian metric of 3-space and $a(t)$ is the scale factor, which we normalize to 1 at current epoch $t_0$: $a(t_0)=1$. Below we put the speed of light equal to 1 ($c=1$), therefore the time has the dimension of a length, the Hubble constant of an inverse length, the mass has dimension of energy, velocity is dimensionless and so on. When Hubble constant appears in the traditional dimension of [km/(s$\cdot$Mpc)] the speed of light appears in [km/s]. 
\subsection{Scalar field models}
The next specification of the scalar field model is used: its equation of state is $P_{de}(a)=w(a)\rho_{de}(a)$ and the temporal derivative of the dark energy pressure is proportional to the temporal derivative of its energy density, $\dot{P}_{de}=c_{a\;(de)}^2\dot{\rho}_{de}$,
where the coefficient $c_{a\;(de)}^2$, often called ``adiabatic sound speed'', is constant. The last condition together with the differential energy-momentum conservation law, which in the Friedmann-Robertson-Walker metric is 
\begin{eqnarray}
a\rho_{de}'=-3(1+w(a))\rho_{de}, \label{rho'}
\end{eqnarray}
gives the ordinary differential equation for $w(a)$:
\begin{eqnarray}
aw'=3(1+w)(w-c_{a\;(de)}^2),\label{w'}
\end{eqnarray}
where a prime denotes the derivative with respect to the scale factor $a$. It has the analytic solution: 
\begin{equation}
w(a)=\frac{(1+w_e)(1+w_0)}{1+w_0-(w_0-w_e)a^{3(1+w_e)}}-1,\label{w} 
\end{equation}
where $c_{a\;(de)}^2$ we have denoted by $w_e$ since it corresponds to the EoS parameter at the beginning of expansion ($w_e=w(0)$), which we call early EoS parameter. One can see also that other constant value here, the integration constant $w_0\equiv w(1)$, is EoS parameter at the current epoch. The differential equation (\ref{rho'}) with (\ref{w}) has the analytic solution for $\rho_{de}$:
\begin{eqnarray}
\rho_{de}(a)=\rho_{de}^{(0)}\frac{(1+w_0)a^{-3(1+w_e)}+w_e-w_0}{1+w_e},\label{rho}
\end{eqnarray}
where $\rho_{de}^{(0)}=3H_0^2\Omega_{de}/8\pi G$ is the dark energy density at current epoch.
It is regular function of $a$ for any values of $w_0$ and $w_e$, excluding singular point $a=0$ for $w_e>-1$. The pressure of fields with parametrization (\ref{w}) is regular function of $a$ too: 
\begin{eqnarray}
 P_{de}(a)=w_e\rho_{de}(a)+(w_0-w_e)\rho_{de}^{(0)},\label{beos}
\end{eqnarray} 
which is the generalized linear barotropic equation of state. That is why we call such models ``the scalar field models of dark energy with barotropic equation of state''.
For the quintessential scalar field ($w_e,\,w_0\,>-1$) both functions, $\rho_{de}(a)$ and $-P_{de}(a)$, decrease monotonously and tend to the asymptotic value $(w_e-w_0)\rho_{de}^{(0)}/(1+w_0)$, the sign of which depends on the ratio of values of $w_e$ and $w_0$.
It means that $P_{de}(a)$ and $\rho_{de}(a)$ can change the sign at different moments of time depending on the relation between $w_e$ and $w_0$. For example, in the case of quintessential scalar field ($w_e,\,w_0>-1$) with $w_0>w_e$ the pressure changes the sign from minus to plus at $a_{(P=0)}=[(1+w_0)w_e/(w_e-w_0)]^{1/3(1+w_e)}$ and the density changes the sign from plus to minus at $a_{(\rho=0)}=[(1+w_0)/(w_e-w_0)]^{1/3(1+w_e)}$. In the opposite case ($w_e>w_0$) their signs ($P<0$, $\rho>0$) always remain unaltered.\\
\indent Note that substituting (\ref{w}) into r.h.s. of (\ref{w'}) we see that the sign of $w'$ is completely determined by the sign of $-(1+w_0)(w_e-w_0)$. 
So, for quintessential field $w(a)$ is monotonously decreasing ($w'<0$) function when $w_e>w_0$. The repulsion property of such field raises with time. When $w_e<w_0$, $w(a)$ is monotonously increasing ($w'>0$) function. The repulsion property of the field in this case recedes with time. For the phantom field, for which $w_0<-1$, the conditions for raising and receding repulsion properties of the field are opposite. In the particular case $w_e=w_0$ the EoS parameter is constant, $w'=0$. Another particular case  $w_e=-1$ or $w_0=-1$ is simply $\Lambda$-term ($w=const=-1$). It must be noted also, that $w(a)$ in the form (\ref{w}) does not allow the phantom divide ($w=-1$) crossing since in this point $\dot{P}_{de}$ and $\dot{\rho}_{de}$ become 0, but it is applicable in both ranges $w>-1$ and $w<-1$ separately.\\
\indent The goal of the paper is analysis of the possibility of distinguishing the scalar fields with receding and raising repulsion by available observational data.\\
\begin{table}
 \caption{The best fitting values and 1$\sigma$
 confidential ranges of cosmological parameters in the
 CSF+CDM model determined by the Markov Chain Monte Carlo technique 
 using the available observational data (for details see \cite{Novosyadlyj2010}). First column -- model of DE with decreasing EoS parameter, second one -- model of DE with increasing EoS parameter. The current Hubble parameter $H_0$ is in units km$\,$s$^{-1}\,
 $Mpc$^{-1}$, the age of the Universe $t_0$ is given 
 in Giga years.}
 \medskip
 \begin{tabular}{c|c|c}
 \hline
 & & \\
 Parameters&$\textbf{p}_1$&$\textbf{p}_2$ \\
 & & \\
 \hline
 & & \\
$\Omega_{de}$&0.72$_{-0.06}^{+0.04}$&0.71$_{-0.05}^{+0.04}$\\
 & & \\
$w_0$& -0.93$_{-0.07}^{+0.13}$&-0.99$_{-0.01}^{+0.16}$\\
 & & \\
$w_e$& -0.97$_{-0.03}^{+0.97}$&-0.05$_{-0.94}^{+0.05}$\\
 & & \\
10$\omega_b$& 0.225$_{-0.013}^{+0.017}$&0.225$_{-0.012}^{+0.017}$\\
 & & \\
$\omega_{cdm}$& 0.111$_{-0.012}^{+0.013}$&0.113$_{-0.013}^{+0.010}$\\
 & & \\
$H_0$& 69.2$_{- 5.1}^{+ 4.2}$&68.6$_{- 4.4}^{+ 4.7}$\\
 & & \\
$n_s$& 0.97$_{-0.03}^{+0.05}$&0.97$_{-0.04}^{+0.05}$\\
 & & \\
$\log(10^{10}A_s)$& 3.07$_{-0.08}^{+0.11}$&3.09$_{-0.09}^{+0.10}$\\
 & & \\
$\tau_{rei}$&0.084$_{-0.037}^{+ 0.049}$&0.090$_{-0.039}^{+0.043}$\\
 & & \\
 \hline
$-\log L$&4027.35&4027.51\\
 \hline
 \end{tabular}
 \label{tabl1}
 \end{table}
\begin{figure}[t]
\includegraphics[width=.47\textwidth]{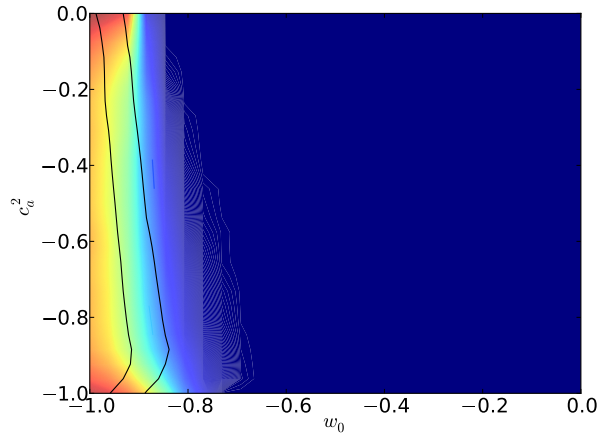}
\caption{Two-dimensional mean likelihood distribution in the plane $w_e-w_0$ for the combined dataset WMAP7+SDSS LRG DR7+SN Union2+HST+BBN. Solid lines correspond to 1$\sigma$ and 2$\sigma$ confidence contours.}
\label{2d_likes}
\end{figure}
\indent The discussed above three parameters of a scalar field -- $\Omega_{de}$, $w_e$ and $w_0$ -- are quite enough for the analysis of background dynamics, luminosity distance -- redshift and angular diameter distance -- redshift relations. But scalar fields are gravitationally unstable and their clustering properties depend also on one more of their characteristic -- the effective sound speed $c_s^2\equiv\delta P_{de}/\delta\rho_{de}$. Indeed, the equations for evolution of Fourier amplitudes of density $\delta_{(de)}\equiv\delta\rho_{de}/\rho_{de}$ and velocity $V_{(de)}$ perturbations of scalar field in the synchronous gauge are 
\begin{eqnarray}
&&\dot{\delta}_{(de)}+3(c_s^2-w)aH\delta_{(de)}+(1+w)\frac{\dot{h}}{2}\nonumber\\ 
&&+(1+w)\left[k+9a^2H^2\frac{c_s^2-w_e}{k}\right]V_{(de)}=0, \label{d_de}\\
&&\dot{V}_{(de)}+aH(1-3c_s^2)V_{(de)}-\frac{c_s^2k}{1+w}\delta_{(de)}=0,\label{V_de}
\end{eqnarray}
where $k$ is the wave number, $h\equiv h_{\alpha}^{\alpha}$ is the trace of scalar perturbations of metric (see e.g. \cite{Novosyadlyj2010,Putter2010} or for gauge-invariant approach e.g. \cite{Novosyadlyj2009,Sergijenko2009b} and citing therein). The equations for the rest of components (non-relativistic and relativistic) are the same as in \cite{Ma1995}. For determination of the fourth parameter of dark energy, $c_s^2$,
we specify the scalar field Lagrangian. We consider the scalar fields with Klein-Gordon (KG) and Dirac-Born-Infeld (DBI) Lagrangians, also called classical and tachyonic respectively,
\begin{eqnarray}
L_{clas}=X-U(\phi),\,\,\,
L_{tach}=-\tilde{U}(\xi)\sqrt{1-2\tilde{X}},\label{L}
\end{eqnarray}
where ${U}(\phi)$ and $\tilde{U}(\xi)$ are the field potentials, $X=\phi_{;i}\phi^{;i}/2$ and $\tilde{X}=\xi_{;i}\xi^{;i}/2$ are kinetic terms. For the homogeneous background
the field variables are connected with the variables of fluid approach by simple relations: 
\begin{eqnarray}
&U(a)=\frac{1-w(a)}{2}\rho_{de}(a),&\,\,\,\tilde{U}(a)=\rho_{de}(a)\sqrt{-w(a)}\label{Uw}\\
&X(a)= \frac{1+w(a)}{2}\rho_{de}(a),&\,\,\,\,\,\,\,\,\tilde{X}(a)=\frac{1+w(a)}{2}.\label{Xw}
\end{eqnarray}
The effective (rest-frame) sound speed $c_{s\,(de)}^2$ for the scalar field with given Lagrangian is defined as
\begin{eqnarray}
c_{s\,(de)}^2\equiv\frac{P_{,X}}{\rho_{,X}}=\frac{L_{,X}}{2XL_{,XX}+L_{,X}}.\label{cs2}
\end{eqnarray}
\begin{figure*}
\includegraphics[width=.47\textwidth]{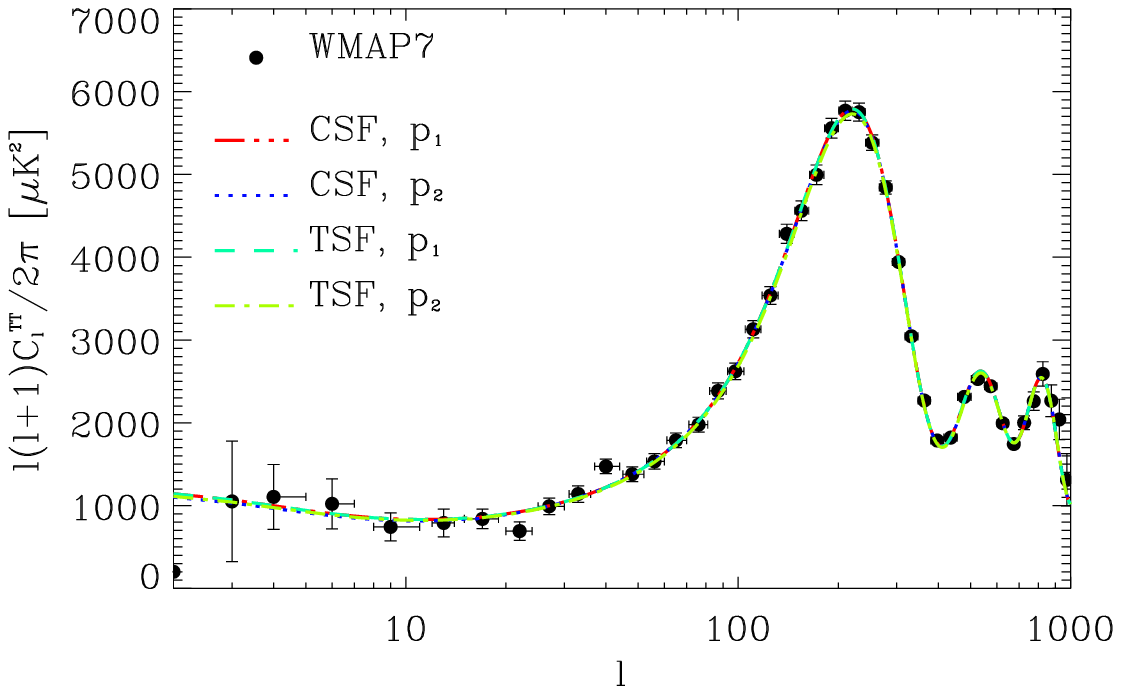}
\includegraphics[width=.47\textwidth]{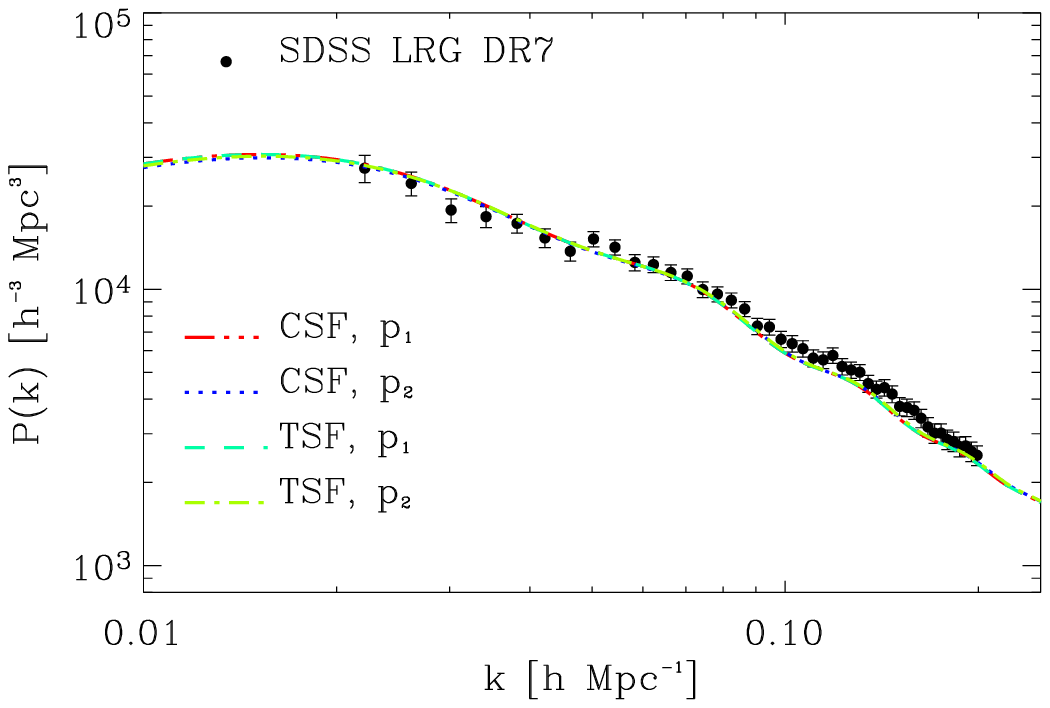}
\caption{The power spectra of CMB temperature fluctuations (left) and matter density ones (right) for cosmological models with classical and tachyonic scalar fields (CSF and TSF respectively) with two sets of the best fitting parameters from Table \ref{tabl1}. The corresponding observational data from WMAP7 and SDSS LRG DR7 are shown by the filled circles.}
\label{cl_pk}
\end{figure*}
It equals 1 for the classical scalar field and $-w$ for the tachyonic one. Therefore, these fields are quite different, can affect the evolution of dark matter perturbations and leave ``fingerprints'' in the large scale structure of the Universe, as it was shown in our previous papers \cite{Novosyadlyj2010,Novosyadlyj2009,Sergijenko2009b}. The goal of this paper is the analysis of possibility of distinguishing the scalar fields with KG and DBI Lagrangians by available observational data.\\
\indent For calculation of the dynamics of expansion of the Universe, the evolution of perturbations in all components, the power spectra of matter density perturbations and CMB anisotropy we have used the publicly available code CAMB \cite{camb,camb_source}, modified to include the presented here expressions (\ref{w}), (\ref{rho}), (\ref{cs2}) and equations (\ref{d_de})-(\ref{V_de}). 

\subsection{Determination of DE and cosmological parameters} 

Therefore, the scalar field model of dark energy, described in the previous subsection, involves three parameters $\Omega_{de}$, $w_e$ and $w_0$ which must be determined by comparison of the calculated predictions on dynamics of expansion and large scale structure of the Universe with corresponding observational data. Since all predictions and data are related with other components (dark matter, baryons, thermal cosmic radiation) it should be done jointly with other cosmological parameter. Concentrating on the analysis of possibility of determination of dark energy parameters we consider the cosmological model with minimal set of six parameters: density parameter of baryons $\omega_{b}$, density parameter of cold dark matter $\omega_{cdm}$, Hubble constant $H_0$, spectral index of initial matter density power spectrum $n_s$, amplitude of initial matter density power spectrum $A_s$ and reionization optical depth $\tau_{rei}$. So, we have nine unknown parameters, but the number of independent ones is eight, since we have assumed the spatial flatness of the Universe. Indeed, the dark energy density parameter $\Omega_{de}$ in this case can be calculated from the zero curvature condition: $\Omega_{de}=1-\Omega_b-\Omega_{cdm}$, where $\Omega_b=\omega_b{\rm h}^2$, $\Omega_{cdm}=\omega_{cdm}{\rm h}^2$, where ${\rm h}\equiv H_0/100{\rm km/s\cdot Mpc}$. \\
\indent To determine the best fitting values and confidential ranges of the  scalar field parameters  together with other cosmological ones in our previous work \cite{Novosyadlyj2010} we have performed the Markov Chain Monte Carlo (MCMC) analysis for the set of current observational data, which include the power spectra from WMAP7 \cite{wmap7a,wmap7b} and SDSS DR7 \cite{Reid2009}, the Hubble constant measurements \cite{Riess2009}, the light curves of SN Ia \cite{SNUnion2} and Big Bang nucleosynthesis (BBN) prior \cite{bbn}.\\ 
\indent We have used the publicly available package CosmoMC \cite{cosmomc,cosmomc_source}, which includes the code CAMB  \cite{camb,camb_source} for calculation of the model predictions for sampled sets of 8 cosmological parameters: $w_e$, $w_0$, $\omega_b$, $\omega_{cdm}$, $H_0$, $n_s$, $A_s$ and  $\tau_{rei}$. The CosmoMC code has been modified to be run with the proposed here parametrization of dark energy EoS parameter (\ref{w}). The flat priors $-1<w_0,\,c_a^2\le 0$ have been used to take into account the quintessential properties of scalar fields with KG and DBI Lagrangians (lower limit) and the constraints following from observational data related to the recombination and nucleosynthesis epochs (upper one).\\
\indent We have performed two MCMC runs for the eight-parametric flat cosmological model with the classical scalar field. Each run had 8 chains and the number of samples in each chain was $\sim200\,000$. For the first run we have used only the mentioned above flat prior for $w_e$. The set of best fitting parameters determined by this run is marked by $\textbf{p}_1$ and presented in the Table \ref{tabl1} together with 1$\sigma$ confidential ranges. All parameters, excluding $w_e$, are well constrained, the one-dimensional posterior and mean likelihood distributions are close and similar to Gaussian (half-Gaussian for $w_0$), confidential ranges are narrow. The EoS parameter at early epoch $w_e$ is essentially unconstrained: its 1$\sigma$ confidence range is wide and coincides practically with the allowed by prior range [-1, 0]. The mean likelihood and posterior are different and the likelihood is bimodal, as it is shown in Fig. \ref{2d_likes}. The first peak is close to -1, another one to 0. The best fitting value of $w_e$ in the set $\textbf{p}_1$ corresponds to the first peak. In this case $w_e<w_0$ which means that the best fitting scalar field model of dark energy has receding repulsion properties ($w'>0$). \\
\indent In order to obtain the best fitting parameters corresponding to the second peak of the likelihood distribution we have performed analogical run with additional condition $w_e>w_0$. The set of best fitting parameters determined by this run is marked by $\textbf{p}_2$ and presented in the Table \ref{tabl1}. Now, the best fitting value of $w_e$ corresponds to the second visible in the upper left corner of Fig. \ref{2d_likes} peak of mean likelihood distribution. In this case $w_e>w_0$ and we can say that the best fitting scalar field model of dark energy has raising repulsion properties ($w'<0$).\\
\indent As we see, large variation of $w_e$ does not change essentially other parameters: each parameter from the set $\textbf{p}_2$ is in the  1$\sigma$ range of the corresponding one from the set $\textbf{p}_1$ and vice versa. The $-\log{L}$'s (last row of the Table \ref{tabl1}) for both sets are very close. The power spectra of matter density perturbations and CMB temperature fluctuations for both sets of the best fitting parameters $\textbf{p}_1$ and $\textbf{p}_2$ are presented in Fig. \ref{cl_pk}. For the same sets of parameters we have calculated also the spectra for tachyonic scalar field models of dark energy. For all four models corresponding lines are superimposed and well match the observational spectra. Therefore, we have double degeneracy: in type of the dynamics of scalar field (receding-raising repulsion properties) and its Lagrangian (classical-tachyonic). Is it possible to distinguish them in principle? It is subject of the next sections.

\section{Dynamics of expansion of the Universe, SN Ia light curves and statefinders}\label{bckgr}

\begin{figure}
\includegraphics[width=.47\textwidth]{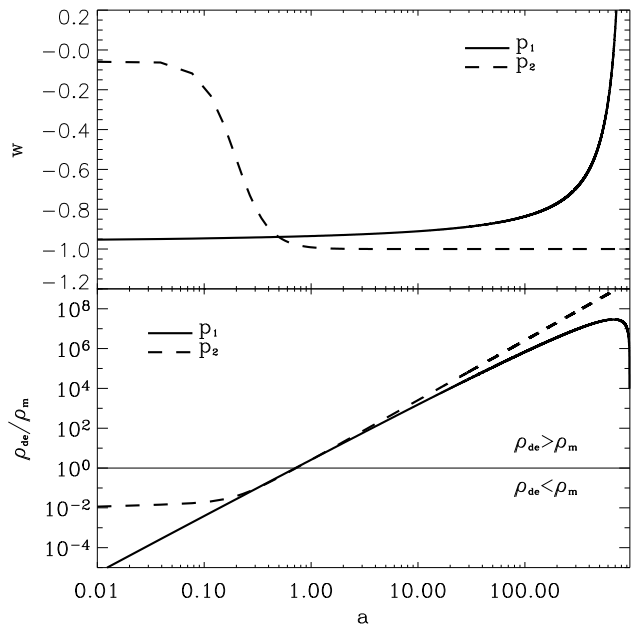}
\caption{The dependences of EoS parameter (top panel) and ratio $\rho_{de}/\rho_{m}$ (bottom panel) on  the scale factor for models with best fitting parameters $\textbf{p}_1$ and $\textbf{p}_2$. The current epoch corresponds to $a=1$.}
\label{fig1}
\end{figure}

\begin{figure*}
\includegraphics[width=.47\textwidth]{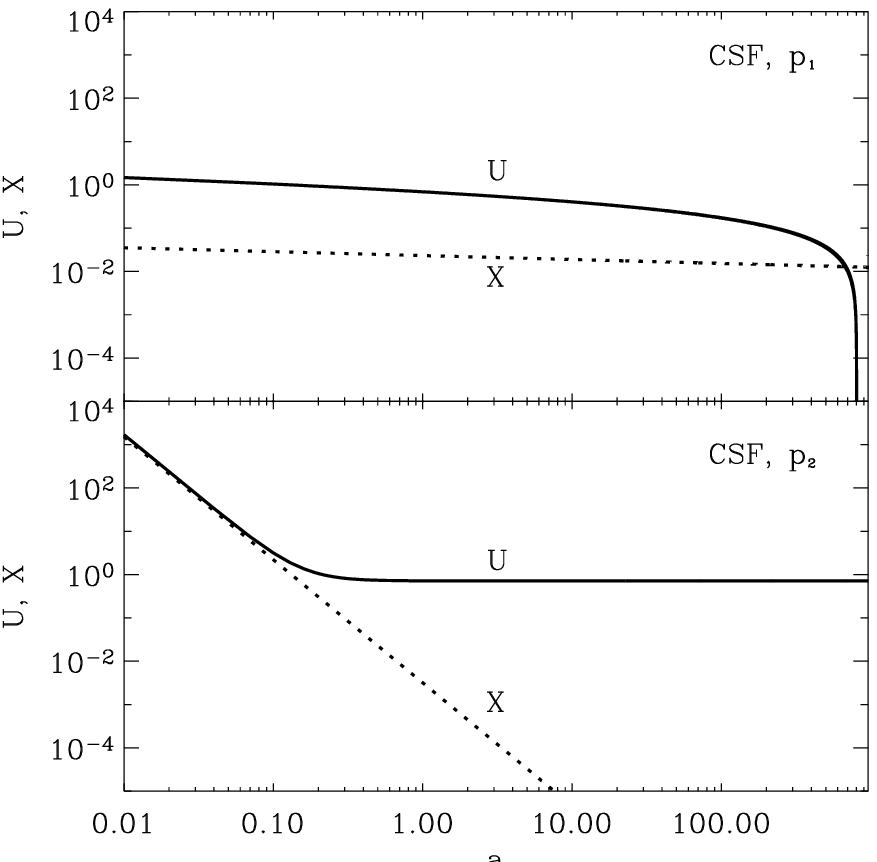}
\includegraphics[width=.47\textwidth]{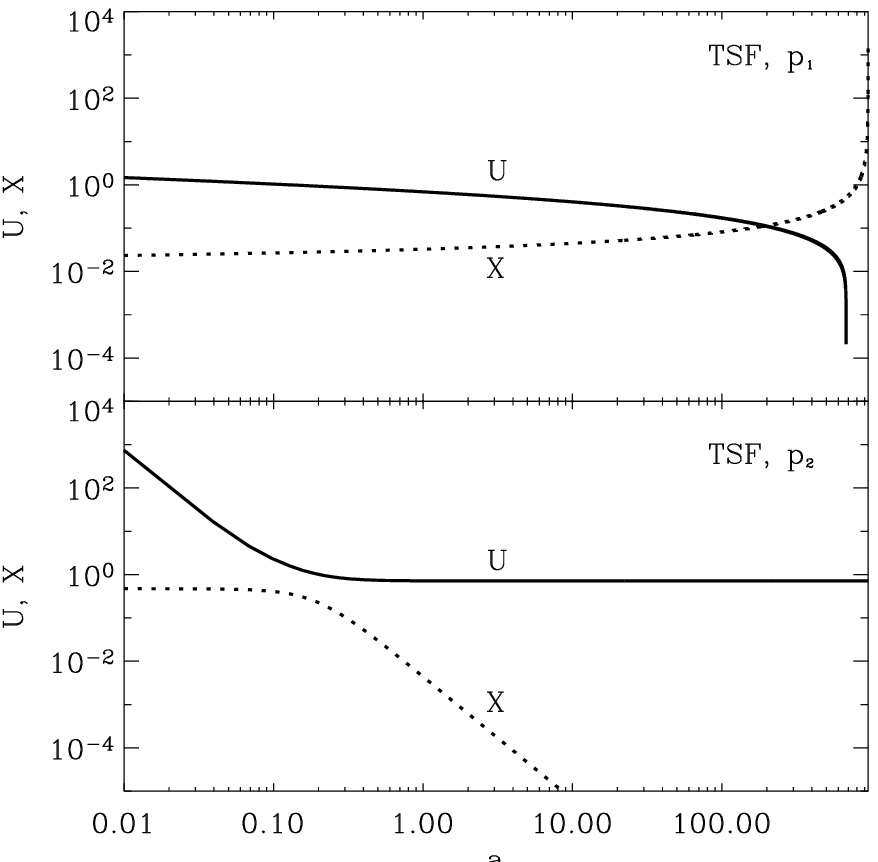}
\caption{Evolution of potentials and kinetic terms of classical (left column) and tachyonic (right column) scalar fields for models with best fitting parameters $\textbf{p}_1$ (top panel) and $\textbf{p}_2$ (bottom panel).}
\label{ux_c}
\end{figure*}

\begin{figure}
\includegraphics[width=.47\textwidth]{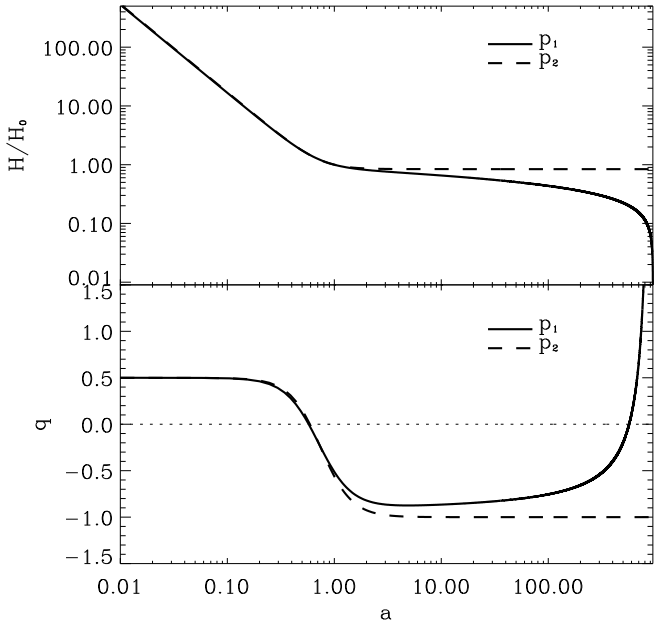}
\caption{The dynamics of expansion of the Universe with barotropic dark energy:  $H(a)$ (top panel) and $q(a)$ (bottom one) for the  $\textbf{p}_1$ and $\textbf{p}_2$ models.}
\label{qhsf}
\end{figure}

\begin{figure}
\includegraphics[width=.47\textwidth]{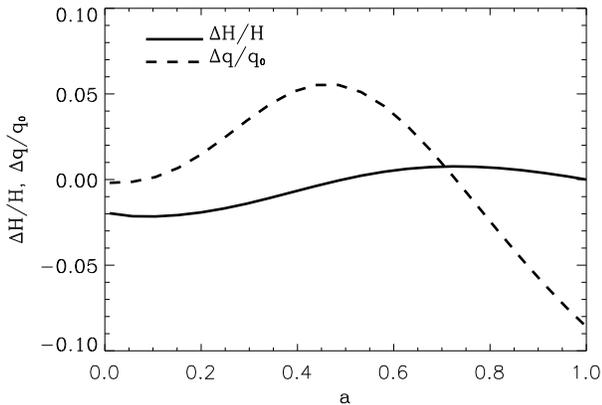}
\caption{The relative differences $\Delta H(a)/H(a)$ and $\Delta q/q_0$ of models with $\textbf{p}_1$ and $\textbf{p}_2$.}
\label{vqhsf}
\end{figure}

\begin{figure}
\includegraphics[width=.47\textwidth]{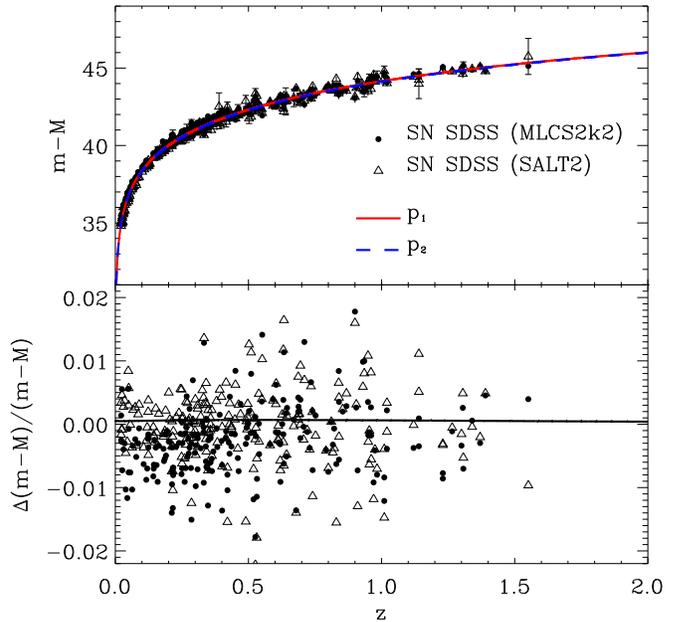}
\caption{Top panel: The distance moduli $(m-M)(z)$ for models with $\textbf{p}_1$ and $\textbf{p}_2$ (lines) and SN SDSS data with different fittings of SN Ia light curves (signs). Bottom panel: the relative difference $\Delta(m-M)/(m-M)$ of two models (line) and scattering of data points from the top panel around the model line with best fitting parameters $\textbf{p}_1$, $\left[(m-M)_{obs}-(m-M)_{mod}\right]/(m-M)_{mod}$.}
\label{dl}
\end{figure}

\begin{figure}
\includegraphics[width=.47\textwidth]{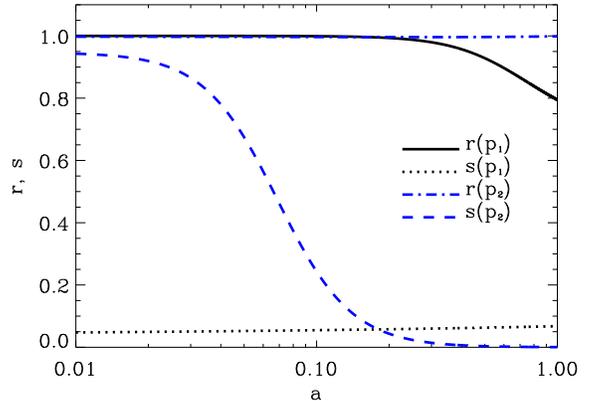}
\caption{Statefinder parameters $r$ and $s$ for models with $\textbf{p}_1$ and $\textbf{p}_2$.}
\label{sr}
\end{figure}

At first we analyze the evolution of scalar field parameters in the homogeneous (background) Universe. The time dependences of the EoS parameter and the ratio of scalar field energy density to dark matter density in the models with best fitting parameters $\textbf{p}_1$ and $\textbf{p}_2$ are presented in Fig. \ref{fig1}. As we see, in spite of closeness of these values for both parameter sets in the current epoch ($a=1$), they are quite different in past ($a<1$) and future ($a>1$). The EoS parameter in the model with $\textbf{p}_1$ increases monotonically from -0.97 at the beginning of expansion to -0.93 at the current epoch and will continue increasing in future up to discontinuity of the second kind, caused by the energy density crossing of zero ($a_{(\rho_{de}=0)}=950$). Immediately after that the Universe reaches the turnaround point ($a_{ta}=a_{(\rho_{de}=0)}+\delta a$, $\delta a/a\ll 1$) and begins to collapse \cite{Novosyadlyj2010}. At the early epoch the scalar field energy density is insignificant similarly to the cosmological constant. In the model with $\textbf{p}_2$ the EoS parameter decreases monotonically from -0.05 at the early epoch to -0.99 now and will asymptotically approach -1 in far future. The energy density in the early epoch traces the density of dark matter. In future it will asymptotically approach the constant value $\rho_{de}^{(\infty)}=\rho_{de}^{(0)}(w_e-w_0)/(1+w_e)$. So, the future of the Universe with such field is de Sitter expansion.\\ 
\indent  The evolution of potentials and kinetic terms of classical and tachyonic scalar fields for models with the best fitting parameters $\textbf{p}_1$ and $\textbf{p}_2$ is shown in the Fig. \ref{ux_c}. One can see, that both fields are quite different for  $\textbf{p}_1$ and $\textbf{p}_2$ parameter sets, but for each of them the potentials of both fields are similar, while kinetic terms are different. In all cases scalar fields roll slowly to the minima of their potentials. In the  models with $\textbf{p}_1$ parameter set it will be reached in finite time ($\approx268$ Gyrs, turnaround point), in the  models with $\textbf{p}_2$ one -- at infinite time. Near the turnaround point in the model with $\textbf{p}_1$ the kinetic terms of both fields become dominating (top panels of Fig. \ref{ux_c}), quintessential scalar fields behave as the k-essential ones.\\
\indent Using the dependences of densities of each component on the scale factor one can deduce from the Einstein equations the following equations for the background dynamics:
\begin{eqnarray}
 H&=&H_0\sqrt{\Omega_r/a^{4}+\Omega_m/a^{3}+\Omega_{de}f(a)}, \label{H}\\
 q&=&\frac{1}{2}\frac{2\Omega_r/a^{4}+\Omega_m/a^{3}+(1+3w)\Omega_{de}f(a)}
{\Omega_r/a^{4}+\Omega_m/a^{3}+\Omega_{de}f_(a)},\label{q}
\end{eqnarray}
where $f(a)=[(1+w_0)a^{-3(1+w_e)}+w_e-w_0]/(1+w_e)$, $\Omega_m=\Omega_b+\Omega_{cdm}$ is the density parameter of matter (nonrelativistic) component at current epoch, $\Omega_r=\Omega_{\gamma}+\Omega_{\nu}$ is the density parameter of relativistic component at current epoch, $H\equiv\dot{a}/{a}$ is the Hubble parameter (expansion rate) for any moment of time and $q\equiv-\ddot{a}/aH^2$ is the acceleration parameter for any moment of time. They describe completely the dynamics of expansion of the homogeneous isotropic Universe.\\
\indent The temporal dependences of $H$ and $q$ for models with $\textbf{p}_1$ and $\textbf{p}_2$ are shown in Fig. \ref{qhsf}. It can be seen that their past evolution ($a\le1$) is practically indistinguishable, but future one is different. In both models the accelerated expansion began at $a\approx 0.58$. At current epoch the acceleration parameters are $q_0\approx-0.57$ in the model with $\textbf{p}_1$  and $q_0\approx-0.52$ in the model with $\textbf{p}_2$. In the last model it will continue decreasing in future, approaching asymptotically -1 as in de Sitter inflation. In the model with $\textbf{p}_1$ the acceleration parameter will reach minimal value $q_{min}\approx-0.87$ at $a_{(q_{min})}\approx4.95$ and then will begin to increase. At $a_{(q=0)}=\left[(1+w_0)(1+3w_e)/(w_e-w_0)/2\right]^{1/3(1+w_e)}\approx570$ the accelerated expansion will be altered by decelerated one. The pressure will become positive at $a_{(w=0)}>\left[w_e(1+w_0)/(w_e-w_0)\right]^{\frac{1}{3(1+w_e)}}\approx678$ and in finite time according to (\ref{beos}) will reach the constant positive value $P_{de}^{(max)}=(w_0-w_e)\rho_{de}^{(0)}\approx 0.035\rho_{de}^{(0)}$ causing the turn around at $a\approx 960$ and collapse. So, the cosmological model with $\textbf{p}_1$ has future finite-time Big Crunch singularity. \\
\indent Fig. \ref{qhsf} illustrates the model degeneracy: two different scalar field models of dark energy are indistinguishable by the dynamics of expansion of the Universe. How deep is this degeneracy? To answer this question we have calculated the relative differences $\Delta H(a)/H(a)$ and $\Delta q/q_0$ of models with $\textbf{p}_1$ and $\textbf{p}_2$. The results are presented in Fig. \ref{vqhsf}. One can see that the appreciable difference ($1\%$) of $H$ in models with $\textbf{p}_1$ and $\textbf{p}_2$ in the past ($a<1$) appears at redshifts $z>1$ ($a<0.5$) and reaches $2\%$ at high $z$. The relative difference of acceleration parameters in these models is maximal ($\approx8\%$) at current epoch ($a\approx1$) and $\approx 5\%$ at the beginning of accelerated expansion ($a\approx0.4-0.6$).\\
\indent Since both values are usually determined from the luminosity distance-redshift relation, it is interesting to compare the difference of distance moduli
\begin{eqnarray}
(m-M)=5\log{d_L}+25, \label{dist_mod}
\end{eqnarray}
where
\begin{eqnarray}
d_L=\frac{c}{H_0}(1+z)\int_0^z\frac{dz'}{\sqrt{\Omega_m(1+z')^3+\Omega_{de}f(\frac{1}{1+z'})}},\nonumber
\end{eqnarray}
for these models with observational uncertainties of their determinations from the light curves of SN. In Fig. \ref{dl} we present the distance moduli for both models and relative difference of them together with corresponding current observational data \cite{Kessler2009}. Both models match observational data on SN SDSS distance moduli  equally good, the difference between them does not exceed 0.1\%, which is essentially lower than dispersion of observational points around of the best fitting curve. To distinguish them other tests based on accurate measurements of acceleration parameter in the vicinity of the Local Group should be proposed or radical improvement of the accuracy of existing ones should be made. \\
\indent If in the future high-precision observations it is possible to measure the second derivative of $H$ then statefinder parameters $r=1+3\dot{H}/H^2+\ddot{H}/H^3$ and $s=(r-1)/[3(q-1/2)]$, introduced by \cite{Sahni2003}, can be used for distinguishing the scalar field models with 
different time variable EoS parameters. For matter plus dark energy dominated epoch they can be presented in our parametrization (\ref{w}) as follows:
\begin{eqnarray*}
r=1+4.5(1+w)w_e\Omega_{de}(a), \quad\quad s=(1+w)w_e/w,
\end{eqnarray*}
where $\Omega_{de}(a)\equiv 8\pi G\rho_{de}(a)/3H^2$.
Their dependences on scale factor $a$ for the best fitting models $\textbf{p}_1$ and $\textbf{p}_2$ are shown in Fig. \ref{sr}. One can see that the differences between $s(\textbf{p}_1)$ and $s(\textbf{p}_2)$ at high $z$ as well as between $r(\textbf{p}_1)$ and $r(\textbf{p}_2)$ at low $z$ are essentially larger than for parameters $H$ and $q$ (Fig. \ref{qhsf} and \ref{vqhsf}).
\section{Matter density and CMB power spectra}\label{spectra}
\begin{figure*}
\includegraphics[width=.47\textwidth]{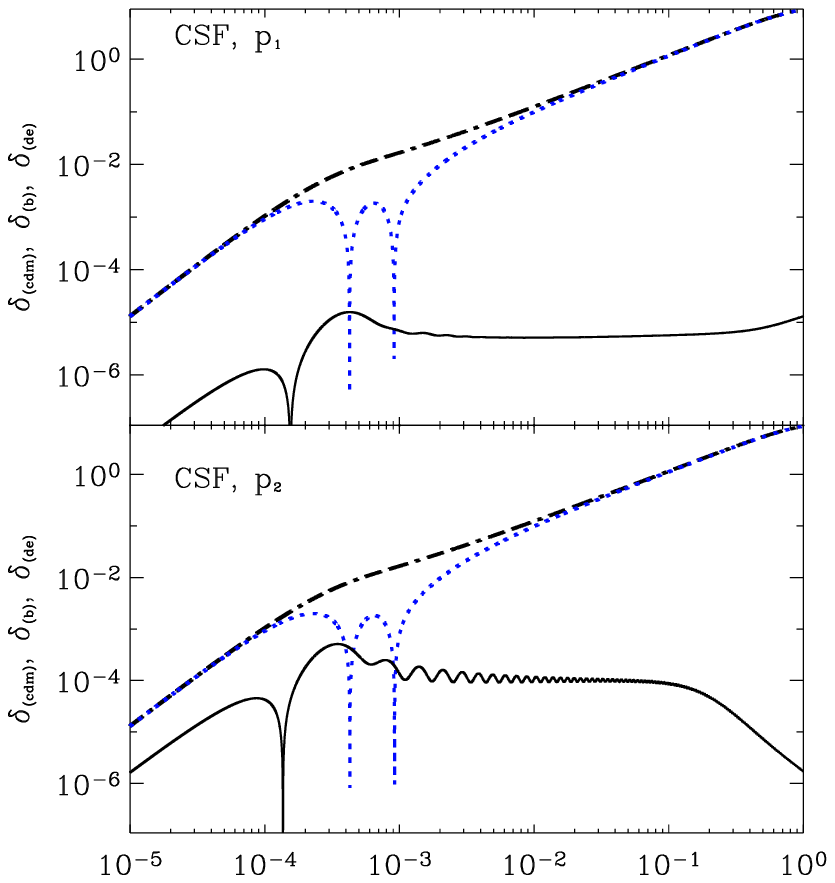}
\includegraphics[width=.47\textwidth]{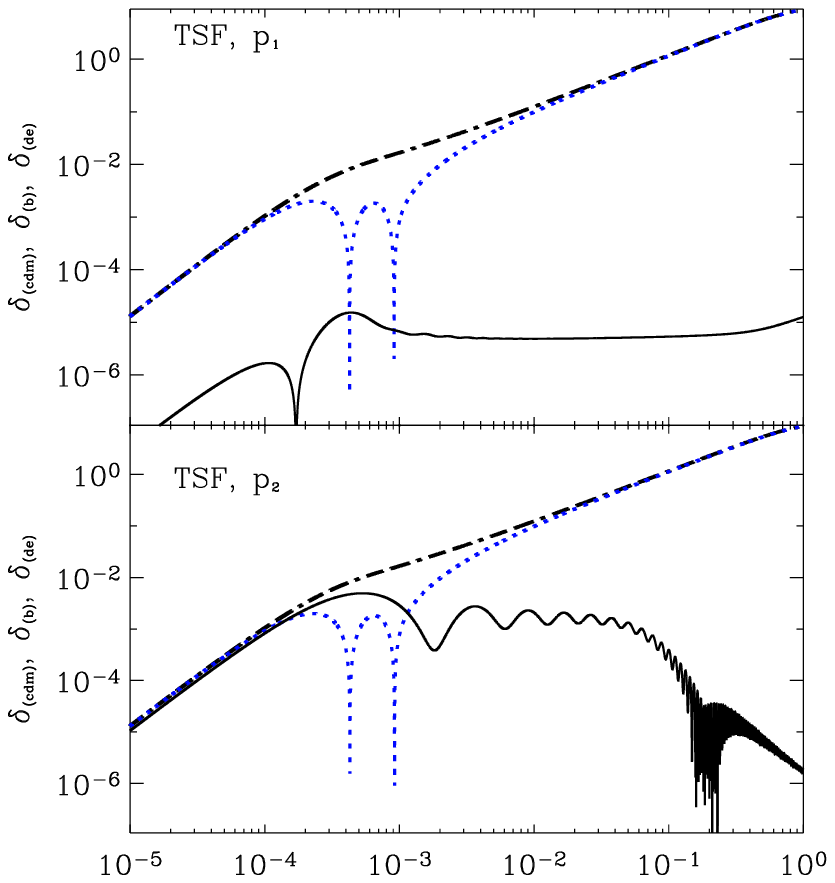}
\caption{Evolution of linear density perturbations of cold dark matter (dash-dotted line), baryons (dotted) and scalar field (solid) for models with $\textbf{p}_1$ and $\textbf{p}_2$. Classical scalar field -- left column, tachyonic one -- right. The wave number of perturbations is $k=0.05$ Mpc$^{-1}$. }
\label{ddeb}
\end{figure*}

\begin{figure*}
\includegraphics[width=.47\textwidth]{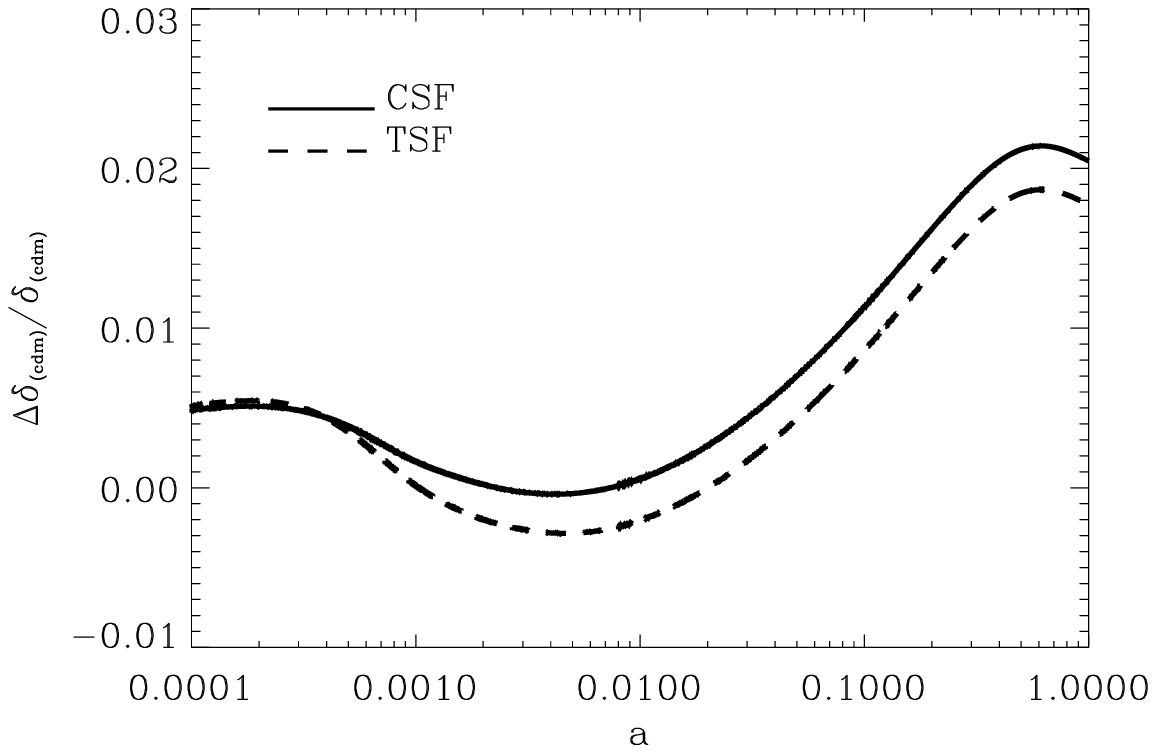}
\includegraphics[width=.47\textwidth]{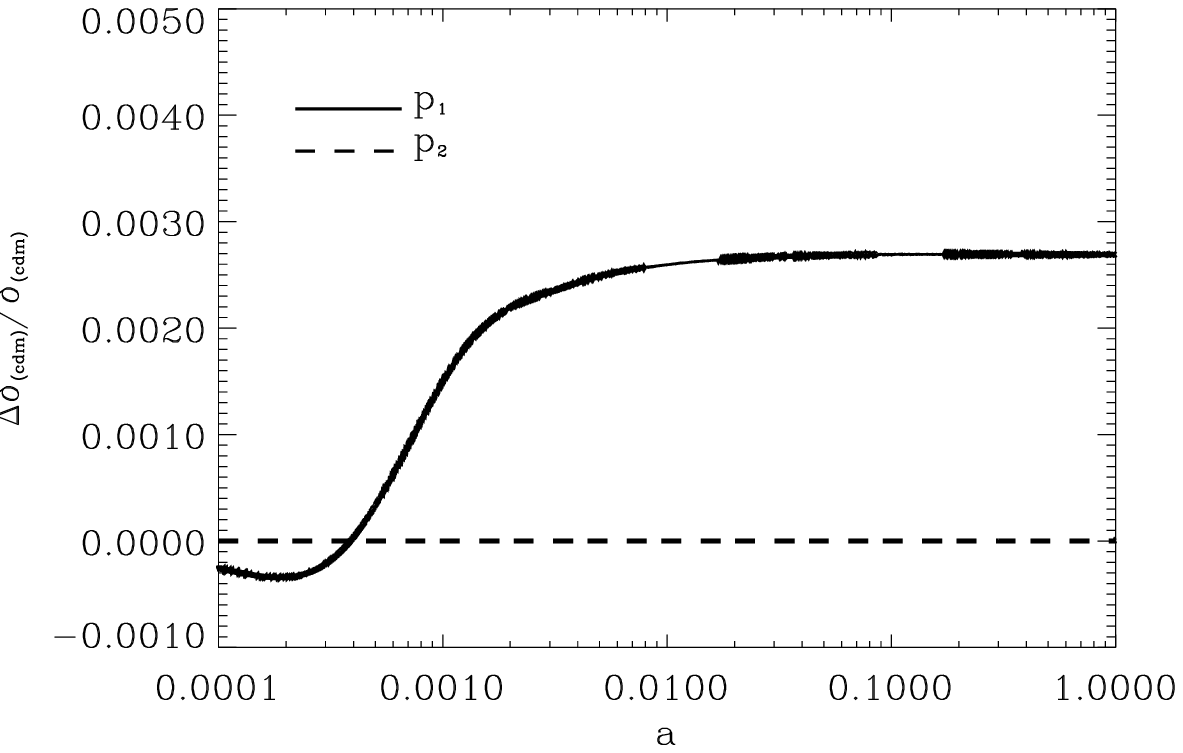}
\caption{The influence of scalar field perturbations on the evolution of linear matter density ones ($k=0.05$ Mpc$^{-1}$). Left panel: the relative difference of CDM density perturbations in models with sets of parameters $\textbf{p}_1$ and $\textbf{p}_2$ $\Delta\delta_{(cdm)}/\delta_{(cdm)}=(\delta_{(cdm)}(\textbf{p}_1;a)- \delta_{(cdm)}(\textbf{p}_2;a))/\delta_{(cdm)}(\textbf{p}_1;a)$ for CSF (solid line) and TSF (dashed line). Right panel: the relative difference of CDM density perturbations in the models with CSF and TSF $\Delta\delta_{(cdm)}/\delta_{(cdm)}=(\delta_{(cdm)}(\textrm{TSF};a)- \delta_{(cdm)}(\textrm{CSF};a))/\delta_{(cdm)}(\textrm{CSF};a)$ for the sets of parameters $\textbf{p}_1$ (solid line) and $\textbf{p}_2$ (dashed line).}
\label{vddeb}
\end{figure*}

\begin{figure*}[t]
\includegraphics[width=.47\textwidth]{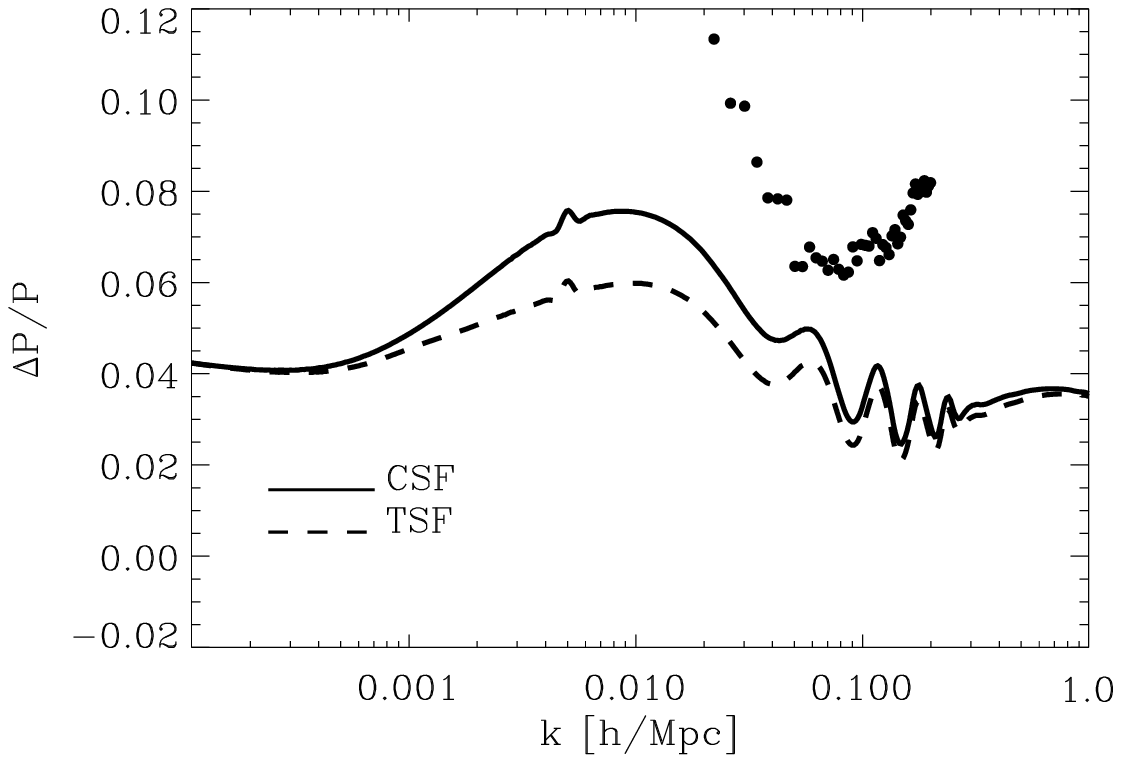}
\includegraphics[width=.47\textwidth]{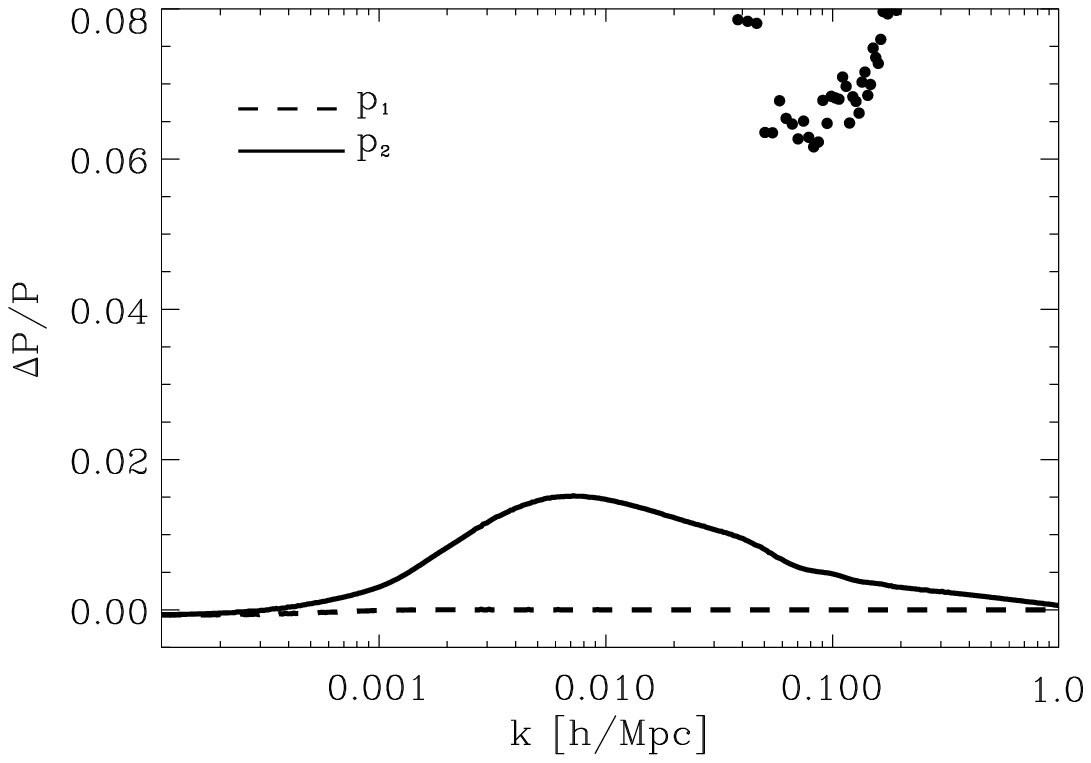}
\caption{Left panel: the relative difference of matter density power spectra $\Delta P/P$ in the models with best fitting parameters $\textbf{p}_1$ and $\textbf{p}_2$ for classical and tachyonic scalar fields. Right panel: the relative difference of matter density power spectra $\Delta P/P$ in the models with classical and tachyonic scalar fields for two sets of the best fitting parameters $\textbf{p}_1$ and $\textbf{p}_2$. Dots show observational uncertainties (1$\sigma$) of SDSS LRG DR7 data.}
\label{vpkm}
\end{figure*}

\begin{figure*}[t]
\includegraphics[width=.47\textwidth]{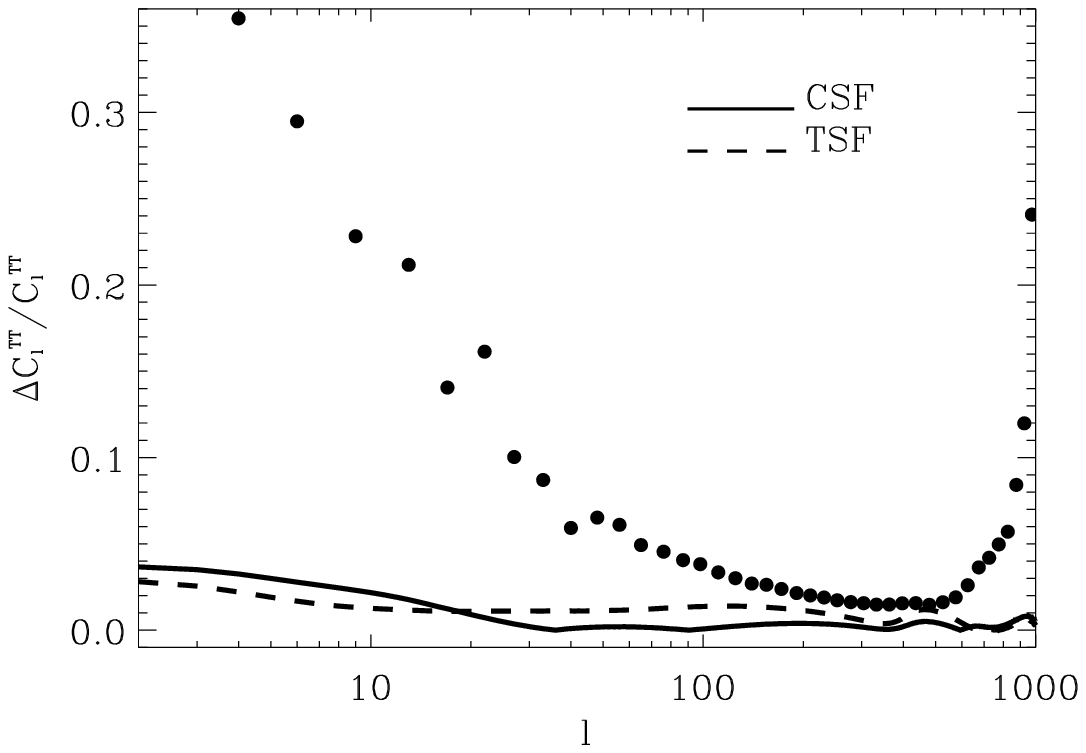}
\includegraphics[width=.47\textwidth]{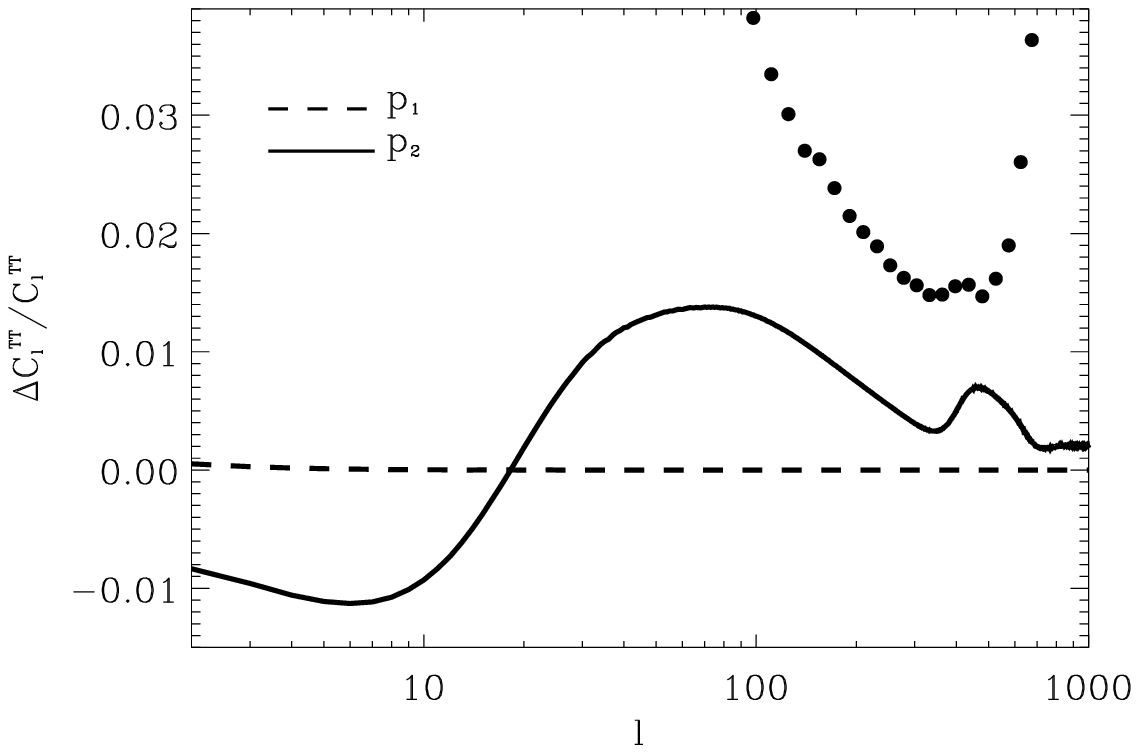}
\caption{Left panel: the relative difference of CMB temperature fluctuations power spectra $\Delta C_{\ell}^{TT}/C_{\ell}^{TT}$ in the models with best fitting parameters $\textbf{p}_1$ and $\textbf{p}_2$ for classical and tachyonic scalar fields. Right panel: the relative difference of CMB temperature fluctuations power spectra $\Delta C_{\ell}^{TT}/C_{\ell}^{TT}$
in the models with classical and tachyonic scalar fields for two sets of the best fitting parameters $\textbf{p}_1$ and $\textbf{p}_2$. Dots show observational uncertainties (1$\sigma$) of WMAP7 data.}
\label{vcl}
\end{figure*}

Let us analyze the large scale structure and CMB anisotropy characteristics calculated for models with $\textbf{p}_1$ and $\textbf{p}_2$ parameter sets, their differences and model distinguishability in the light of them. The dark energy affect these characteristics via dynamics of expansion of the homogeneous Universe (scale-independent growth factor) and gravitational interaction between perturbations of all components for the same $k$-mode. The evolution of scalar field density perturbations depends also on 
$c_s^2$, therefore the influence of them on the matter density ones will be different for KG and DBI Lagrangians. Using the modified CAMB code we have integrated the Einstein-Boltzmann system of equations for all components (scalar field, dark matter, baryons, thermal radiation, massless active neutrinos) in the cosmological models with parameter sets $\textbf{p}_1$ and $\textbf{p}_2$. The results for baryons, cold dark matter and scalar fields with KG/DBI Lagrangians (wave number of perturbations $k=0.05$ Mpc$^{-1}$) are shown in Fig. \ref{ddeb}. In the case of model with $\textbf{p}_1$ the evolution tracks of scalar field density perturbations are similar for both Lagrangians, but they differ in the case of model with $\textbf{p}_2$. Since scalar field density perturbations are not observable, let us look how they influence the evolution tracks of matter density perturbations. For this purpose in Fig. \ref{vddeb} we present the differences $\Delta\delta_{(cdm)}/\delta_{(cdm)}=\left(\delta_{(cdm)}(\textbf{p}_1;a)-\delta_{(cdm)}(\textbf{p}_2;a)\right)/\delta_{(cdm)}(\textbf{p}_1;a)$ for KG and DBI scalar field models (left panel) as well as $\Delta\delta_{(cdm)}/\delta_{(cdm)}=\left(\delta_{(cdm)}(\textrm{CSF};a)-\delta_{(cdm)}(\textrm{TSF};a)\right)/\delta_{(cdm)}(\textrm{CSF};a)$ for $\textbf{p}_1$ and $\textbf{p}_2$ parameter sets (right panel). The differences between models with $\textbf{p}_1$ and $\textbf{p}_2$ for scalar fields with KG and DBI Lagrangians reach $\sim 2\%$ at the current epoch. In the case of $\textbf{p}_1$ parameter set the scalar field with DBI Lagrangian is indistinguishable from the scalar field with KG one (dashed line in right panel of Fig. \ref{vddeb}). In the case of $\textbf{p}_2$ parameter set the difference between both fields does not exceed 0.3\%. It means that at the current level of accuracy of observational data and numerical codes for calculations of model predictions they cannot be distinguished too. For final conclusion other $k$-modes should be analyzed.\\
\indent For such purpose we have calculated by the modified version of CAMB the power spectra of matter density perturbations for the same four models $P(CSF,\textbf{p}_1;k)$, $P(CSF,\textbf{p}_2;k)$, $P(TSF,\textbf{p}_1;k)$ and $P(TSF,\textbf{p}_2;k)$ (right panel of  Fig. \ref{cl_pk}). The relative differences 
$\Delta P/P$ are presented for them in Fig. \ref{vpkm}. The observational relative 1$\sigma$ errors of SDSS LRG DR7 data \cite{Reid2009} are shown there too. The maximal differences between spectra in models with $\textbf{p}_1$ and $\textbf{p}_2$ are at $k\sim0.01$ h/Mpc ($\approx 8$\% in the models with CSF and $\approx 6$\% in the models with TSF). At lower scales, $k\sim0.1$ h/Mpc, they are $\sim4-5\%$, that is lower than the observational uncertainties at this scales. In the case of $\textbf{p}_1$ parameter set the models with CSF and TSF are indistinguishable, the relative difference is less than 0.1\%.  In the case of $\textbf{p}_2$ parameter set it is maximal ($\approx 1.5$\%) at $k\sim$0.01 h/Mpc, but it is $\approx 0.6$\% at $k\sim$0.1 h/Mpc, where observational uncertainties are lowest, $\approx 6$\%. So, the possibility of distinguishing CSF from TSF by $P$ is unpredictable. And vice versa, distinguishing of the model with $\textbf{p}_1$ from one with $\textbf{p}_2$ by the matter density power spectrum data expected in upcoming decade looks possible.\\ 
\indent Let us analyze now the possibility of distinguishing these four models by observational data on CMB anisotropy. In Fig. \ref{cl_pk} the angular power spectra of CMB temperature fluctuations $C_{\ell}^{TT}$ \cite{tb4} for them are presented in left panel. All four lines are visually superimposed and go near most points in their 1$\sigma$ error bars. The relative differences between them $\Delta C_{\ell}^{TT}/C_{\ell}^{TT}$ as well as relative 1$\sigma$ errors of WMAP7 data \cite{wmap7a,wmap7b} are shown in Fig. \ref{vcl}. The maximal differences ($\sim3-4\%$) between $C_{\ell}^{TT}$ of models with $\textbf{p}_1$ and $\textbf{p}_2$ are at low spherical harmonics, where cosmic variance is too large to distinguish between such models. In the range of acoustic peaks $\ell\sim 100-700$, where observational data are most accurate ($\sim 2\%$), the difference between $C_{\ell}^{TT}$ of models with $\textbf{p}_1$ and $\textbf{p}_2$ is somewhat smaller ($\le1.5\%$). So, the expected data releases on CMB anisotropy from WMAP and Planck teams would probably allow the possibility to answer the question ''which scalar field dark energy, with decreasing or increasing EoS parameter, fills our Universe?''. To answer the question ''which is the field Lagrangian?'' it will be harder in the case of decreasing EoS parameter and practically impossible in the case of increasing one (see right panels of Figs. \ref{vddeb}-\ref{vcl}).\\ 
\indent To support the latter conclusions let us compare the relative differences of CMB temperature fluctuations $\Delta C_{\ell}^{TT}/C_{\ell}^{TT}$ and polarization $\Delta C_{\ell}^{EE}/C_{\ell}^{EE}$  power spectra in models with both fields and parameter sets with the observational uncertainties, modelled for the Planck satellite in the following way. Assuming that the noise part is due to the combined effect of Gaussian beam and spatially uniform Gaussian white noise, for the experiment with known beam width and sensitivity the noise power spectrum for each channel can be approximated as follows:
\begin{eqnarray*}
N_{\ell}^{jj}=\theta^2_{fwhm}\sigma_j^2\exp\left[\ell(\ell+1)\frac{\theta^2_{fwhm}}{8\log2}\right],
\end{eqnarray*}
where $j$ stands for either $TT$ or $EE$, $\theta_{fwhm}$ is the full width at half maximum of the Gaussian beam and $\sigma_j$ is the root mean square of the instrumental noise. The non-diagonal noise terms vanish since the noise contributions from different maps do not correlate. For the experiments with more than one channel the total noise power spectrum is obtained as:
\begin{eqnarray*}
\frac{1}{N_{\ell}^{jj(tot)}}=\sum_{i=1}^{n_{chan}}\frac{1}{N_{\ell}^{jj(i)}},
\end{eqnarray*}
where $n_{chan}$ is the number of channels. The described procedure was proposed by \cite{futurcmb} and implemented in their code FuturCMB \cite{futurcmb_source}, which we use here.
 
\begin{figure*}
\includegraphics[width=0.95\textwidth]{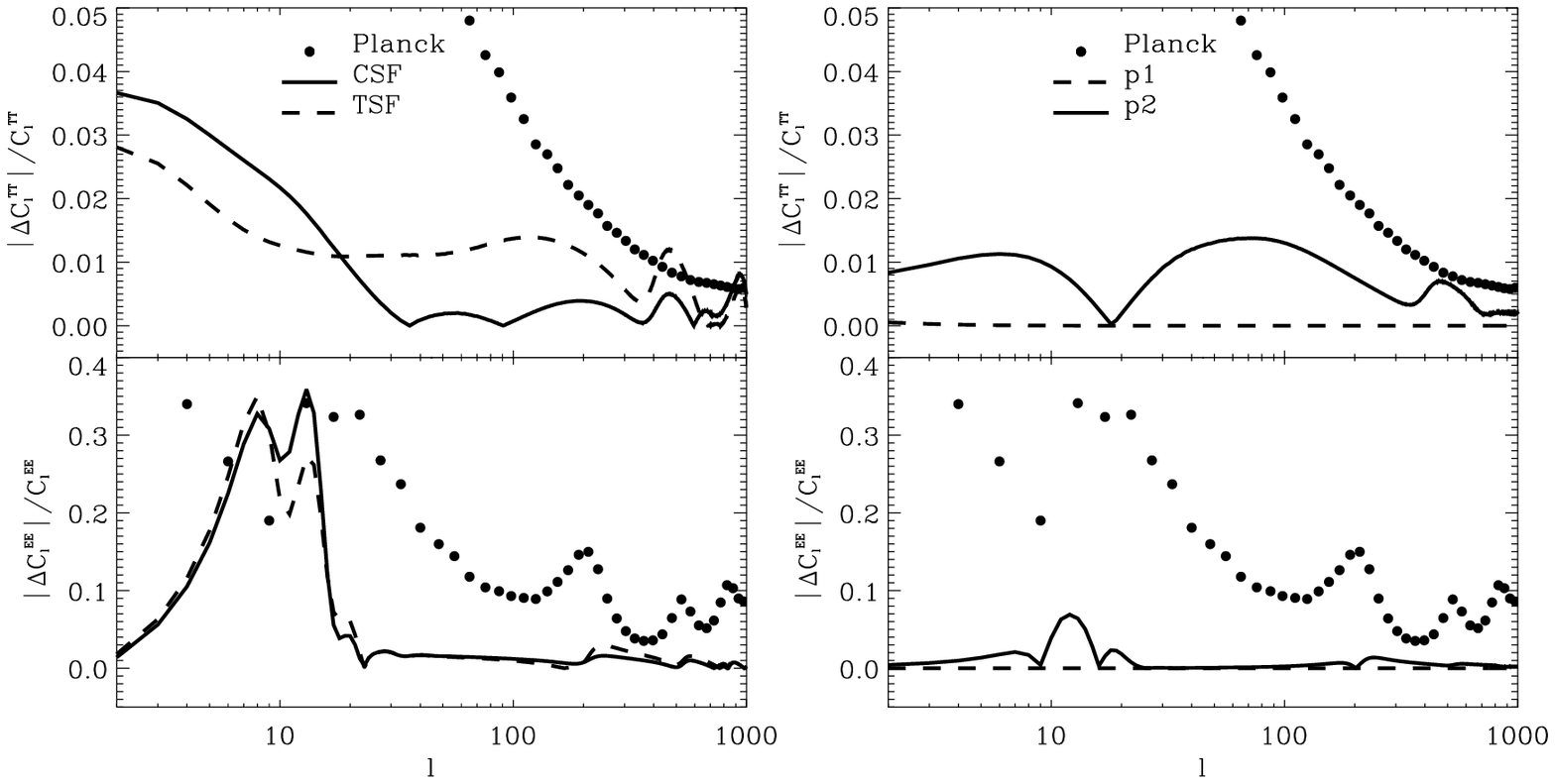}
\caption{Left: the relative differences of CMB temperature fluctuations $\Delta C_{\ell}^{TT}/C_{\ell}^{TT}$ and polarization $\Delta C_{\ell}^{EE}/C_{\ell}^{EE}$ power spectra (from top to bottom) in the models with classical and tachyonic scalar fields for two sets of the best fitting parameters $\textbf{p}_1$ and $\textbf{p}_2$. Right: the relative differences of CMB temperature fluctuations $\Delta C_{\ell}^{TT}/C_{\ell}^{TT}$ and polarization $\Delta C_{\ell}^{EE}/C_{\ell}^{EE}$ power spectra (from top to bottom) in the models with two sets of the best fitting parameters $\textbf{p}_1$ and $\textbf{p}_2$ for either classical or tachyonic scalar field. Dots show the observational uncertainties, modelled for the Planck satellite.}\label{planck}
\end{figure*}

In Fig. \ref{planck} we show the estimated errors for the Planck experiment with 3 channels (for each of them $\theta_{fwhm}$, $\sigma_T$ and $\sigma_E$ are 9.5 arcmin, 6.8 $\mu K$ per pixel and 10.9 $\mu K$ per pixel; 7.1 arcmin, 6.0 $\mu K$ per pixel and 11.4 $\mu K$ per pixel; 5.0 arcmin, 13.1 $\mu K$ per pixel and 26.7 $\mu K$ per pixel correspondingly). The observed sky fraction is assumed to be $f_{sky}=0.65$.  The models with the same fields but different parameter sets $\textbf{p}_1$ and $\textbf{p}_2$ can be distinguished by the data with such precision. For the CMB temperature fluctuations power spectrum the difference between studied models exceeds the estimated error level at high spherical harmonics, while for the polarization power spectrum at low spherical harmonics, where it is maximal. The models with different fields but the same parameter sets (corresponding to both decreasing and increasing EoS parameters) are still indistinguishable at such level of experimental precision.

\section{Discussion and conclusion}\label{conclusions}

Weakness of constraints on dark energy parameters for models with time variable EoS is widely discussed in the literature \cite{Corasaniti2004,Pogosian2005,Hollenstein2009}. 
Moreover, the degeneracy between them limits strongly the possibility to test whether $w$ is constant or not. Our results, presented in the Tab. 2 of \cite{Novosyadlyj2010}, confirm that. We have used the set of observational data including the light curves of SN Ia, the power spectra from WMAP7 and SDSS DR7, the Hubble constant measurements and BBN prior to analyze the possibility of determining the parameters of dynamical scalar field jointly with other cosmological ones. Processing of the MCMC chains for eight-parametric models ($\textbf{p}=(\Omega_{de},\,w_0,\,w_e,\,\omega_{b},\,\omega_{cdm},\,H_0,\,n_s,\,A_s,\,\tau_{rei})$) gives marginalized posterior and likelihood distributions as well as the best fitting values of parameters and their confidential ranges. For the most parameters the posterior and likelihood distributions are unimodal and similar, the confidential ranges are narrow. The exception is $w_e$, for which the marginalized posterior distribution is unimodal, but marginalized mean likelihood distribution is bimodal. The values of maximal likelihood for both peaks are very close. The first peak corresponds to the scalar field model of dark energy with increasing EoS parameter, the second to model with decreasing one. The question is: which accuracy of observational data is necessary to distinguish these two models of dark energy? We have shown, that the difference of luminosity distance moduli $\Delta(m-M)$ in the models with decreasing and increasing EoS parameter does not exceed 0.1\% (Fig. \ref{dl}), while the most accurate observational data on SN luminosity distances (SN SDSS, MLCS2k2 and SALT2 \cite{Kessler2009}) disperse around the best fitting model line in the range of $\sim 2\%$. Therefore, the radical improvement of statistical and systematic uncertainties of data is necessary for distinguishing between these models of dark energy. Then the statefinder parameters $r$ and $s$ \cite{Sahni2003} allow the possibility to differentiate effectively the scalar field models of dark energy with increasing and decreasing EoS parameters.\\
\indent To diminish or remove the degeneracy between dark energy, dark matter and curvature parameters other data must be also improved. The difference of the matter power spectra is maximal ($\sim$6-8\%) at scales 0.1-0.5 $h^{-1}$Gpc (left panel of Fig. \ref{vpkm}). The maximal accuracy of matter density perturbations power spectra obtained in the current sky surveys reaches $\sim$6\% at scales $\sim 50\,h^{-1}$Mpc (SDSS DR7), which is somewhat above the difference between model power spectra at this scales. If expected data from the next releases are 1.5-2 times more accurate, then they allow us to distinguish surely the models with decreasing and increasing EoS parameters. The important data in the used set are the data on CMB anisotropy, as the difference of angular power spectra of CMB temperature fluctuations is $\sim$1-4\%, which is somewhat smaller than accuracy of the current measurements by WMAP at scales $\sim 0^o.5-2^o$ ($\ell\sim 100-500)$ (left panel of Fig. \ref{vcl}). We believe that the expected data release from WMAP team (nine-year observations) and particularly expected Planck data together with other improved data will make it possible to determine whether the EoS parameter of dark energy is increasing, decreasing or constant.\\
\indent Possibility of distinguishing the scalar field models of dark energy with different Lagrangians is less optimistic. We have analyzed only two from many proposed in the literature ones and found that differences between predictions of such models are essentially lower than accuracy of current observational data. So, the scalar fields with KG and DBI Lagrangians are indistinguishable at the current level of accuracy of observational data (right panels of Fig. \ref{vddeb}-\ref{vcl}). Only in the case of decreasing EoS parameter the scalar fields with KG and DBI Lagrangians are potentially distinguishable, perhaps already by Planck data. But in the case of increasing EoS parameter they are practically indistinguishable, since differences for them are $\ll0.1\%$, which is essentially lower than the accuracy of theoretical and computer calculations of predictions. These small differences have the simple explanation: in such case the densities of scalar fields were too low in the past and did not leave the fingerprints in the CMB anisotropy and LSS. This conclusion can be generalized for any type of scalar fields - quintessence, phantom, quintom, etc.\\
\indent We have considered the principal possibility of distinguishing the scalar fields with decreasing and increasing EoS parameters and two Lagrangians in cosmological models with the minimal set of parameters. Adding extra parameters -- masses of active neutrinos, sterile neutrino, tensor mode, primordial magnetic fields, etc. -- weakens constraints on dark energy parameters and possibility to distinguish the different scalar fields modelling it. So, only join progress of theory, laboratory experiments, astrophysical and cosmological observations will lead to the unveiling of nature of dark energy.

\begin{acknowledgments}
This work was supported by the project of Ministry of Education and Science of Ukraine (state registration number 0110U001385), research program ``Cosmomicrophysics'' of the National Academy of
Sciences of Ukraine (state registration number 0109U003207) and the SCOPES project No. IZ73Z0128040 of Swiss National Science Foundation. Authors also acknowledge the usage of CAMB and FuturCMB packages.

\end{acknowledgments}

\end{document}